# Stability of stagnation via an expanding accretion shock wave

A. L. Velikovich,[1] M. Murakami,[2] B. D. Taylor,[3] J. L. Giuliani,[1] S. T. Zalesak,[4] and Y. Iwamoto[5]

[1]*Plasma Physics Division, Naval Research Laboratory, Washington, DC 20375, USA*
[2]*Institute of Laser Engineering, Osaka University, Osaka 565-0871, Japan*
[3]*Laboratory for Computational Physics & Fluid Dynamics, Naval Research Laboratory, Washington, DC 20375, USA*
[4]*Berkeley Research Associates, Beltsville, MD 20705, USA*
[5]*Ehime University, Matsuyama, Ehime Pref. 790-8577, Japan*

Stagnation of a cold plasma streaming to the center or axis of symmetry via an expanding accretion shock wave is ubiquitous in inertial confinement fusion (ICF) and high-energy-density plasma physics, the examples ranging from plasma flows in x-ray-generating Z pinches [Y. Maron *et al.,* Phys. Rev. Lett. **111**, 035001 (2013)] to the experiments in support of the recently suggested concept of impact ignition in ICF [H. Azechi *et al*., Phys. Rev. Lett. **102**, 235002 (2009); M. Murakami *et al*., Nucl. Fusion **54**, 054007 (2014)]. Some experimental evidence indicates that stagnation via an expanding shock wave is stable, but its stability has never been studied theoretically. We present such analysis for the stagnation that does not involve a rarefaction wave behind the expanding shock front and is described by the classic ideal-gas Noh solution in spherical and cylindrical geometry. In either case the stagnated flow has been demonstrated to be stable, initial perturbations exhibiting a power-law, oscillatory or monotonic, decay with time for all the eigenmodes. This conclusion has been supported by our simulations done both on a Cartesian grid and on a curvilinear grid in spherical coordinates. Dispersion equation determining the eigenvalues of the problem and explicit formulas for the eigenfunction profiles corresponding to these eigenvalues are presented, making it possible to use the theory for hydro code verification in two and three dimensions.

I. INTRODUCTION

Supersonic stagnation of a cold plasma streaming to the center or axis of symmetry is a key element of many laser- and magnetically-driven inertial confinement fusion (ICF) and high-energy-density physics (HEDP) experiments, see Refs. 1-13 and references therein. In both indirect- and direct-drive laser fusion[1-3] such stagnation constitutes the first stage of the hot spot formation, between the convergence of the leading shock wave in the vapor at the target center and the moment when the reflected shock wave reaches the dense shell.[4] The recently proposed impact-ignition approach to the ICF[5] uses thermalization of the kinetic energy of a laser-accelerated plasma in an accretion shock wave as the main mechanism of plasma heating, thereby requiring much higher implosion velocities, ~1000 km/s, than conventional laser fusion.[1-3] Experimental demonstration of plasma heating to fusion temperatures in such stagnation has been reported for spherical[6] and planar[7] geometry.

In cylindrical geometry, thermalization of the kinetic energy of magnetically-driven plasma in an accretion shock wave is a prominent feature of Z-pinch implosions.[8,9] The kinetic-to-thermal energy conversion via an expanding cylindrical accretion shock wave has been demonstrated to play a major role in the keV x-ray radiation production in both gas-puff and wire-array Z-pinches.[10] The same is likely to apply to the deuterium gas-puff Z-pinch implosions on the Z facility that produced large yields of DD fusion neutrons.[11] It has also been demonstrated that the early-time ablation of wires in cylindrical wire-array Z-pinch implosions produces an almost uniform, low-density, cold precursor plasma that coasts to the pinch axis, stagnating via an expanding accretion shock wave into a fairly stable precursor plasma column, see Refs. 12, 13 and references therein.

The term "accretion shock wave" originally comes from astrophysics, where it denotes the shock front separating the cold infalling gas from the dense stagnated plasma in a variety of stellar objects, from white dwarfs to core-collapse supernovae, see Refs. 14, 15 and references therein. Under the astrophysical conditions, the star gravity plays a major role in accelerating the infalling gas towards the accretion shock. Under the ICF/HEDP experimental conditions the mechanism of plasma stagnation via the expanding accretion shock is the same, although the gravity does not play a role in its acceleration. Still, there is a sufficient similarity to justify the recently proposed laboratory-astrophysics experiments[16] aimed at observing the so-called standing accretion shock instability[17] in a laser-driven plasma stagnating via a hemispherical shock wave.

A planar stagnation via an accretion shock wave, like that demonstrated in Ref. 7, is known to be stable. Indeed, such a plasma flow is essentially the same as a shock-piston flow driven by a rigid piston moving at a constant velocity. An isolated planar shock wave is stable unless the equation of state of the shocked material has some peculiarities, see §90 of Ref. 18. In particular, it is always stable for an ideal gas with any value of its constant adiabatic exponent $\gamma$. The same is true for a planar shock-piston flow.[19,20] Stability of an expanding shock flow, however, is a more complex issue. The best-known examples of such a flow are spherical and cylindrical blast waves described by the famous Sedov self-similar point-blast solution.[21] An expanding blast wave in an ideal gas had been theoretically predicted to exhibit an oscillatory power-law instability growth if the gas $\gamma$ was sufficiently small.[22,23] Such instability has indeed been observed in laser-driven experiments with strongly radiating gases,[24] see also Refs. 25, 26 and references therein.

The main physical difference between the planar shock-piston and the expanding blast-wave flows is that the shock wave is supported in the former case and unsupported in the latter. Only the energy initially released in the point blast is available to drive the blast wave. Being stable *per se* for any $\gamma$, the shock wave produced by the point blast is immediately followed by a rarefaction wave that gradually slows it down. It has been argued[27] that the Vishniac instability of a blast wave[22,23] is caused by the shock interaction with the strong oscillations in the downstream rarefaction part of the blast-wave flow. Such strong oscillations in an unsupported shock wave have been theoretically predicted[28] and experimentally observed[29] in a planar geometry.

Stagnation via an expanding accretion shock wave may be more similar to a planar shock-piston flow than to an expanding blast-wave flow. The shock wave representing the boundary of the stagnated plasma is supported by the kinetic energy of the incident plasma. It might slow down as it propagates but it does not have to. The simplest and the best-known example of such stagnating flow is the one-dimensional (1D) self-similar Noh solution,[30] which has been used for verification of practically every code designed to model implosions, explosions and propagation of shock waves. It describes a stagnation of a converging flow of cold plasma in a constant-velocity shock wave that converts all the kinetic energy of the incident plasma into the thermal energy of the resting, uniform stagnated plasma. In the absence of a rarefaction wave behind the expanding shock wave, there seems to be no physical reason for a hydro instability development, and one can expect the stagnating flow to be as stable as the shock front itself.

The long experience of successful verification of two- and three-dimensional (2D and 3D) hydrocodes against the classic 1D Noh solution is an implicit confirmation, a sign – although not a proof – of the hydrodynamic stability of stagnation via an expanding shock wave. There is also some experimental evidence supporting stability of the accretion-shock flows in HEDP. Observed dense precursor plasma column is remarkably symmetric and stable,[12,13] in contrast with the stagnated plasma column produced after the implosion of the main wire-array mass, which in most cases is strongly affected by the magnetic Rayleigh-Taylor instability. Still, as far as we know, a small-amplitude stability analysis of the plasma stagnation in an expanding accretion shock wave has not been done even for the classic Noh solution. Note that since publication of the original Noh's work, a few generalizations of his solution, herein referred to as classic, have been obtained, both in ideal gas dynamics (see the online supporting material to Ref. 31) and in MHD.[32,33]

Stability analysis of the classic Noh solution is of interest for several reasons. First, the stability problem is solved analytically, so its explicit solution could be used for verification of hydrocodes in two and three dimensions. Second, this is a necessary first step in the stability analysis of more general and realistic stagnating accretion-shock flows described by the available generalizations of the classic Noh solution. In the extended family of ideal-gas 1D self-similar solutions describing stagnation in an expanding accretion shock wave, to which the classic Noh solution belongs, the most important one in this respect probably is the Guderley reflected-shock solution,[34-36] which closely approximates the early stage of the hot-spot formation in laser fusion targets.[1-4] This solution involves a partly supported shock wave that, similarly to one produced by a point blast, is slowed down from behind by a rarefaction wave. Therefore it is not clear in advance whether or not it becomes unstable for certain values of gas $\gamma$ and mode number $l$, just like the converging-shock Guderley solution is.[37] Stability analysis of the reflected-shock Guderley case, will be a natural continuation of both this work and Ref. 37, where such analysis has been done for a converging shock wave. Stability analysis of other hydrodynamic[31] and MHD[32,33] generalizations of the classic Noh solution could shed light on the stability of hot-spot formation and other cases of shock-wave stagnation in laser-fusion and Z-pinch implosions, in laboratory astrophysics experiments,[16] as well as guide verification of hydrodynamic and MHD codes.

Our linear, small-amplitude stability analysis covers the general case of 3D perturbations of the classic Noh solution for spherical geometry, with small-amplitude distortion of the expanding shock front proportional to the spherical harmonic, $Y_l^m(\theta,\phi)$, and a special case of filamentation 2D perturbations $\sim \exp(im\phi)$ for cylindrical geometry. For the reasons explained below, the general case of 3D perturbations for cylindrical geometry, with the distortion of the expanding shock proportional to $\exp(im\phi+ikz)$ needs to be studied separately. For the above two cases our perturbation problem is solved analytically, resulting in an explicit dispersion equation for the eigenvalues determining the time evolution of the solutions, as well as explicit formulas for the corresponding eigenfunctions. For both spherical and cylindrical cases, the stagnation via a constant-velocity expanding accretion shock wave turns out to be stable. For all mode numbers, for any value of the gas adiabatic exponent $\gamma$, the distortion amplitude of the expanding shock front decreases as a power of time. In most cases such a decrease is oscillatory, but in some special cases it could be monotonic.

This paper is structured as follows. In Section II we describe the formulation and analytic solution of our perturbation problem (all the derivations in full detail are given in the online Supplemental material, Ref. 38). In Section III we present numerical simulations of the unperturbed classic Noh problem on a Cartesian grid in two and three dimensions, and of the Noh problem with a small initial two-dimensional perturbation added to the solution obtained on a curvilinear grid in spherical coordinates. We discuss the evolution of the small perturbations indicating the difference between the 1D theoretical solution and the 2D/3D numerical solutions. In Section IV we conclude with a discussion.

II. THEORY

We solve the equations of ideal fluid dynamics

$$\frac{\partial \rho}{\partial t} + \nabla \cdot (\rho \mathbf{v}) = 0, \quad (1)$$

$$\frac{\partial \mathbf{v}}{\partial t} + \mathbf{v}\nabla\mathbf{v} + \frac{1}{\rho}\nabla p = 0, \quad (2)$$

$$\left(\frac{\partial}{\partial t} + \mathbf{v}\nabla\right)\ln\left(\frac{p}{\rho^\gamma}\right) = 0, \quad (3)$$

where $\rho(\mathbf{r},t)$, $p(\mathbf{r},t)$ and $\mathbf{v}(\mathbf{r},t)$ are the density, the pressure, and the velocity, respectively, as functions of the Eulerian coordinate $\mathbf{r}$ and the time $t$. In Eq. (3) the adiabatic relation of an ideal gas is postulated in terms of constant adiabatic exponent $\gamma$.

The classic 1D Noh solution[30] is specified by its initial conditions:

$$\begin{aligned} \rho(r,0) &= \rho_0, \\ p(r,0) &= 0, \\ \mathbf{v}(r,0) &= -v_0 \mathbf{e}_r, \end{aligned} \quad (4)$$

where $\rho_0$ is the initial density, $v_0 > 0$ is the uniform initial radial velocity, $\mathbf{e}_r$ is a unit vector in the positive radial direction. These initial conditions describe an infinite space filled with a uniform cold gas whose initial velocity has the same absolute value $v_0$ everywhere, and it is directed at each point to the plane, axis or center of symmetry in the cases of planar, cylindrical and spherical symmetry, respectively. These cases are denoted by the values of the parameter $\nu = 1$, 2, and 3, respectively.

The 1D expanding-shock flow emerging after $t = 0^+$ maintains its initial planar, cylindrical, or spherical symmetry.[30] Profiles of the density and pressure behind the expanding shock front are uniform, and the gas velocity is zero:

$$\begin{aligned}\rho(r,t) &= \rho_s, \\ p(r,t) &= p_s, \\ v(r,t) &= 0, \quad 0 < r \leq R_s(t),\end{aligned} \qquad (5)$$

where the expanding shock radius is $R_s(t) = v_s t$ and the shock velocity is

$$v_s = \frac{\gamma - 1}{2} v_0; \qquad (6)$$

the velocity of the shock front with respect to the pre-shock gas is $D = v_s + v_0 = (\gamma + 1) v_0 / 2$. The uniform post-shock density and pressure are, respectively,

$$\rho_s = \left(\frac{\gamma + 1}{\gamma - 1}\right)^\nu \rho_0, \qquad (7)$$

$$p_s = \frac{\gamma - 1}{2} \rho_s v_0^2 = \frac{2}{\gamma - 1} \rho_s v_s^2. \qquad (8)$$

The pre-shock profiles are

$$\begin{aligned}\rho(r,t) &= \left(1 + \frac{2}{\gamma - 1} \frac{v_s t}{r}\right)^{\nu - 1} \rho_0, \\ p(r,t) &= 0, \\ v(r,t) &= -v_0 \mathbf{e}_r, \quad R_s(t) < r.\end{aligned} \qquad (9)$$

Obviously, far from the expanding shock front, at $r / v_s t \to \infty$, $\rho(r,t) \to \rho_0$, and the solution (9) stays close to the initial conditions (4). The above classic Noh solution is clearly self-similar: all the fluid variables depend on the radial coordinate $r$ only via the dimensionless combination

$$\xi = \frac{r}{R_s(t)} = \frac{r}{v_s t}, \tag{10}$$

the self-similar coordinate. The time-dependent position of the shock front corresponds to the constant value of $\xi = 1$. Inspecting (9) we observe a major difference between the planar case of $\nu = 1$, on the one hand, and the cylindrical and spherical cases $\nu = 2$, 3, on the other. In the planar case there is no pre-shock convergence of the cold gas, it maintains its initial uniform density before being shocked. In the cylindrical and spherical cases, the initial velocity profile, which remains uniform in the absence of pressure gradient in the cold pre-shock gas, is not compatible with a uniform density profile at $t > 0$ because of convergence of the gas moving at uniform velocity towards the axis or the center. The first line of Eq. (9) describes the corresponding pre-shock density increase for $\nu = 2$, 3. The planar Noh solution is therefore completely equivalent to the shock-piston problem, in which a rigid planar piston at $t = 0$ starts moving with a constant velocity into a half-space filled with a cold uniform gas. Small-amplitude stability analysis of the planar shock-piston problem has been done by many authors, the theory is now comprehensive, see[19,20,39] and references therein, which is why the planar case $\nu = 1$ will not be considered here.

We will study the general case of 3D spherical perturbation problem, where each perturbation mode is characterized by two integer angular mode numbers, $l$ and $m$, so that the corresponding displacement of the shock front is proportional to the spherical function of the polar angle $\theta$ and azimuthal angle $\phi$, $Y_l^m(\theta, \phi) = P_l^m(\cos\theta) \exp(im\phi)$, where $P_l^m(\cos\theta)$ is the associated Legendre function (below we omit the arguments $\theta$ and $\phi$ of $Y_l^m$). This symmetry makes it possible to separate the time and spatial variables in the perturbation problem, as in,[23,37] and to obtain its explicit analytical solution. As for the cylindrical case, the general perturbation eigenmodes are characterized by two parameters, the azimuthal mode number $m$ and the axial wavenumber $k$, the eigenfunctions being proportional to $\exp(im\phi + ikz)$. Since the axial wavenumber is dimensional, associated with a time-independent axial wavelength $\lambda = 2\pi/k$, the general perturbation problem contains two length scales, $k^{-1}$ and $R_s(t)$, one of which is time-dependent. Therefore the time variable cannot be separated, as in Ref. 37; perturbation amplitudes can be complicated functions of the dimensionless time variable $kR_s(t) = kv_s t$, exactly as in the planar case.[19,20,39] Here we limit ourselves to the so-called filamentation modes[40] in cylindrical geometry, assuming $k = 0$, $m \geq 0$. For these eigenmodes the perturbation technique used in Refs. 23 and 37 is fully applicable. The general cylindrical case of finite $k$ and $m$, here, as well as in the stability analysis of a blast wave,[23] needs to be addressed separately.

We assume the cold plasma ahead of the expanding shock front to be unperturbed. The perturbations are nonzero only behind the shock front, which is itself distorted. Expanding its displacement in terms of spherical harmonics, we express the time- and angle-dependent radius of the perturbed spherical shock front as

$$r_s(\theta, \phi, t) = v_s t \left[ 1 + \varepsilon \sum_{l,m} \left( \frac{t}{t_0} \right)^{\sigma_{l,m}} \zeta_1^{l,m} Y_l^m \right], \tag{11}$$

where the dimensionless smallness parameter $\varepsilon \ll 1$ is characteristic of all the first-order perturbation quantities, $\zeta_1^{l,m}$ is a complex dimensionless amplitude of the $(l,m)$ perturbation mode of order unity, $\sigma_{l,m}$ is a complex dimensionless eigenvalue to be determined for this mode, and $t_0$ is a dimensional time unit. The particular choice of the latter value is not relevant, since any other choice is easily accommodated by a scaling transformation of $\varepsilon$.

Similarly, perturbed density and pressure are presented as functions of all three spatial coordinates and time as

$$\rho(r,\theta,\phi,t) = \rho_s \left[ 1 + \varepsilon \sum_{l,m} \left( \frac{t}{t_0} \right)^{\sigma_{l,m}} G_1^{l,m}(\xi) Y_l^m \right], \tag{12}$$

$$p(r,\theta,\phi,t) = p_s \left[ 1 + \varepsilon \sum_{l,m} \left( \frac{t}{t_0} \right)^{\sigma_{l,m}} P_1^{l,m}(\xi) Y_l^m \right]. \tag{13}$$

Here $G_1^{l,m}$ and $P_1^{l,m}$ are dimensionless radial eigenfunctions corresponding to the $(l,m)$ perturbation mode. It is convenient to seek them as functions of the self-similar coordinate (10) because the boundary conditions for these functions are imposed at the shock front, i. e. $\xi = 1$; these functions are therefore sought in the interval $0 \leq \xi \leq 1$.

The perturbed post-shock fluid velocity, which is itself of the first order of smallness, has all three spatial components and it can be decomposed into the sum of its radial and transverse components: $\mathbf{v} = \mathbf{v}_r + \mathbf{v}_\perp = v_r \mathbf{e}_r + \mathbf{v}_\perp$. The radial velocity amplitude has the same angular dependence as the density and pressure perturbations:

$$v_r(r,\theta,\phi,t) = \varepsilon v_s \sum_{l,m} \left( \frac{t}{t_0} \right)^{\sigma_{l,m}} V_{r1}^{l,m}(\xi) Y_l^m, \tag{14}$$

where $V_{r1}^{l,m}$ is the corresponding dimensionless eigenfunction. In the following analysis the transverse velocity $\mathbf{v}_\perp$ is not used in its explicit form. Instead, we calculate its transverse divergence defined by

$$r \nabla_\perp \cdot \mathbf{v}_\perp (r,\theta,\phi,t) = \varepsilon v_s \sum_{l,m} \left( \frac{t}{t_0} \right)^{\sigma_{l,m}} D_1^{l,m}(\xi) Y_l^m, \tag{15}$$

where $\nabla_\perp$ is the transverse divergence operator, and $D_1^{l,m}$ is the corresponding dimensionless eigenfunction, as done in Ref. 37. The scalar amplitude in the left-hand side of (15) is seen to have the same angular dependence as the right-hand sides of Eqs. (11)-(14). As in,[37] the equations and boundary conditions defining our perturbation problem can be fully expressed in terms of the functions $G_1^{l,m}$, $P_1^{l,m}$, $V_{r1}^{l,m}$ and $D_1^{l,m}$. In our case of cylindrical geometry with $k = 0$, one only needs to replace in the right-hand sides of Eqs. (11)-(15) the spherical function $Y_l^m$ with the exponent $\exp(im\phi)$ and to remove the superscript $l$ indicating the polar mode number. Below we will omit the summation symbol and the angular mode numbers $l$ and $m$ in the super- and subscripts of all the eigenfunctions and the eigenvalue $\sigma_{l,m}$, with the understanding that we solve the problem for each particular $(l,m)$ (or $m$, in the cylindrical case) eigenmode separately. For this mode, from now on we assume $\zeta_1^{l,m} = 1$ or $\zeta_1^m = 1$. Although our notation is close to that of Ref. 37, note that the definitions (13)-(15) of the perturbation amplitudes $P_1$, $V_{r1}$ and $D_1$ are not exactly the same as used there.

Substituting (12)-(15) into the fluid equations (1)-(3), in the first order in $\varepsilon$ we obtain our linearized perturbation equations and boundary conditions. They are presented below in the form applicable to both spherical and cylindrical geometries. We introduce the main mode number $j \geq 0$, a non-negative integer. For spherical geometry $j = l$, the polar mode number, and $\nu = 3$. For cylindrical geometry $j = m$, the azimuthal mode number, and $\nu = 2$.

$$\left(\sigma - \xi \frac{d}{d\xi}\right) G_1 + \frac{dV_{r1}}{d\xi} + \frac{1}{\xi}\left[(\nu - 1)V_{r1} + D_1\right] = 0, \tag{16}$$

$$\left(\sigma - \xi \frac{d}{d\xi}\right) V_{r1} + \frac{2}{\gamma - 1}\frac{dP_1}{d\xi} = 0, \tag{17}$$

$$\left(\sigma - \xi \frac{d}{d\xi}\right) D_1 - \frac{2j(j+\nu-2) P_1}{(\gamma-1)\xi} = 0, \tag{18}$$

$$\left(\sigma - \xi \frac{d}{d\xi}\right)(P_1 - \gamma G_1) = 0. \tag{19}$$

One can derive these equations from Eqs. (17)-(20) of Ref. 37, taking into account the difference in notation and the background flow profiles. It is probably easier to derive them directly, as done in Section A of Ref. 38.

The corresponding boundary conditions at the perturbed shock front, $\xi = 1$, are reduced to

$$G_1(1) = -\frac{2(\nu-1)}{\gamma+1}, \tag{20}$$

$$P_1(1) = \frac{2}{\gamma+1}\left[(\gamma-1)\sigma + \gamma - \nu\right], \tag{21}$$

$$V_{r1}(1) = \frac{2(1+\sigma)}{\gamma+1}, \tag{22}$$

$$D_1(1) = \frac{2j(j+\nu-2)}{\gamma-1}, \tag{23}$$

see the derivation in Section B of Ref. 38.

Expressing $V_{r1}$, $D_1$, $G_1$, and their derivatives via $P_1$ with the aid of Eqs. (17), (18), and (19), respectively, we transform Eq. (16) into a second-order equation for $P_1$ that can be reduced to the Gauss hypergeometric equation. We introduce two auxiliary functions

$$f_{\pm}(\xi) = {}_2F_1\left(\frac{j-\sigma}{2}, \frac{j\pm 1-\sigma}{2}; j+\frac{\nu}{2}; M_2^2 \xi^2\right), \tag{24}$$

where $_2F_1(a, b; c; z)$ is the Gauss hypergeometric function, $M_2 = \sqrt{(\gamma-1)/(2\gamma)} < 1$ is the downstream Mach number characteristic of the strong shock wave. The values of these functions at the shock front are denoted by $f_{s\pm} = f_\pm(1)$.

Equations (16)-(19) and the boundary conditions (20)-(23) are satisfied by the following self-similar perturbation profiles (a detailed derivation is found in Sections A and C of Ref. 38). The pressure perturbation profile is given by

$$P_1 = \frac{2[(\gamma-1)\sigma+\gamma-\nu]}{(\gamma+1)f_{s+}} \xi^j f_+(\xi), \qquad (25)$$

Obviously (25) satisfies the boundary condition (21) at $\xi = 1$ identically, i. e., for any $\sigma$. The pressure perturbation (25) describes a sonic wave reverberating in the shocked gas. The density perturbation profile

$$G_1 = \frac{2[(\gamma-1)\sigma+\gamma-\nu]}{\gamma(\gamma+1)f_{s+}} \xi^j f_+(\xi) - \frac{2(\gamma-1)(\sigma+\nu)}{\gamma(\gamma+1)} \xi^\sigma \qquad (26)$$

identically satisfies the boundary condition (20). The first term in the right-hand side of (26) is the contribution to the density perturbation from the sonic wave (25), whereas the second term describes the amplitude of the density/entropy perturbation localized in a fluid particle. Similarly, the transverse velocity divergence profile

$$D_1 = -\frac{4j(j+\nu-2)[(\gamma-1)\sigma+\gamma-\nu]}{(\gamma^2-1)(j-1-\sigma)f_{s+}} \xi^{j-1} f_-(\xi) \\ + \frac{2j(j+\nu-2)}{\gamma-1}\left\{1+\frac{2[(\gamma-1)\sigma+\gamma-\nu]f_{s-}}{(\gamma+1)(j-1-\sigma)f_{s+}}\right\}\xi^\sigma \qquad (27)$$

identically satisfies the boundary condition (23). The first term in the right-hand side of (27) is due to the contribution to the transverse divergence of the curl-free velocity perturbation from the sonic wave, whereas the second term describes the contribution from the divergence-free vortical velocity perturbation localized in a fluid particle. Finally, the radial velocity profile is

$$V_{r1} = \frac{4\left[(\gamma-1)\sigma+\gamma-\nu\right]}{(\gamma^2-1)f_{s+}} \xi^{j-1} \left[f_+(\xi) + \frac{\sigma+1}{j-1-\sigma} f_-(\xi)\right]$$
$$- \frac{2j(j+\nu-2)}{(\sigma+\nu-1)(\gamma-1)} \left\{1 + \frac{2\left[(\gamma-1)\sigma+\gamma-\nu\right]f_{s-}}{(j-\sigma-1)(\gamma+1)f_{s+}}\right\} \xi^\sigma .$$
(28)

The first two terms in the right-hand side of (28) are the contributions to the radial velocity perturbation from the sonic wave, whereas the third term describes the contribution from the vortical velocity perturbation.

Substituting (25)-(28) into Eqs. (16)-(19), one can check that these equations are also satisfied identically, for any $\sigma$. The eigenvalue $\sigma$ is to be determined from the requirement that the radial velocity profile (28) satisfies the boundary condition (22) at $\xi=1$. The latter requirement yields the following dispersion equation for $\sigma$:

$$\left\{(\gamma-1)\sigma^2 + \left[\nu(\gamma-3)+2\right]\sigma - (\gamma+1)j(j+\nu-2) + (\nu-1)(\gamma+1-2\nu)\right\}f_{s+}$$
$$-2\left[(\gamma-1)\sigma+\gamma-\nu\right](\sigma+j+\nu-1)f_{s-} = 0 .$$
(29)

Note that the dispersion equation (29) for spherical geometry, $j=l$, $\nu=3$, does not contain the azimuthal mode number $m$. It turns out that the spectrum of eigenvalues $\sigma$ found from (29) for given mode numbers $(l,m)$ is fully determined by the polar mode number $l$ and independent of the azimuthal mode number $m$, although the corresponding eigenfunctions are explicitly dependent on $m$ via the spherical functions $Y_l^m$. For cylindrical geometry, $j=m$, $\nu=2$, and the azimuthal mode number $m$ determines the perturbation development.

The transcendental dispersion equation (29) in general is not analytically solvable, although for some particular and limiting cases its explicit solutions could be found. As we will demonstrate below, the number of discrete complex eigenvalues is infinite both for the spherical and cylindrical cases. We calculate numerically some of the lowest (i. e., closest to the origin on the complex plane) eigenvalues and discuss the asymptotic properties of the eigenvalues far from the origin.

Note that the positive argument of the hypergeometric functions in (29) is less than ½ for any $\gamma$, which implies a fast convergence of the Gauss hypergeometric series. Calculating partial sums of the series truncated at some high order $N \gg 1$, we reduce the exact dispersion equation to an approximate one, the left-hand side of which is a polynomial in $\sigma$ of the order $2N+2$. All the complex roots of this polynomial are easy to calculate numerically for any $N$. Some of the approximate roots found this way converge as the order of truncation $N$ is increased, and these are identified with the eigenvalues of the original problem. To calculate the eigenvalues shown below, we have checked the numerical convergence up to the order of $N = 45$. Some lowest-order eigenvalues for $\gamma = 5/3$ are tabulated in the Table for spherical and cylindrical geometry. More eigenvalues for several values of $\gamma$ are presented in Tables I and II of Ref. 38 for spherical and cylindrical geometry, respectively. Typical spectra are illustrated in Fig. 1 for low values of $l = 0$, 1 and 2 (spherical) and $m = 0$, 1 and 2 (cylindrical). Since all the coefficients in dispersion equation (29) are real, the eigenvalues $\sigma$ are either real or pairs of complex conjugates. In Fig. 1 and in all our tables we only present the eigenvalues with a non-negative imaginary part.

Table. Lowest-order (radial mode number $n = 1$ to 4) eigenvalues found from Eq. (29) for cylindrical and spherical geometry, $\gamma = 5/3$, and angular mode numbers $m$ and $l$ varied from 0 to 4.

| Mode # | $n$ | Eigenvalues | | | |
|---|---|---|---|---|---|
|  |  | 1 | 2 | 3 | 4 |
| $m$ |  | Cylindrical | | | |
| 0 |  | $-1$ | $-2$ | $-5.148 + 2.328i$ | $-3.546 + 10.61i$ |
| 1 |  | $-2$ | $-4.775$ | $-4.418 + 6.830i$ | $-3.674 + 13.90i$ |
| 2 |  | $-3.506 + 4.254i$ | $-4.970 + 10.43i$ | $-3.952 + 16.92i$ | $-3.709 + 23.73i$ |
| 3 |  | $-3.493 + 7.065i$ | $-5.327 + 13.85i$ | $-4.313 + 19.82i$ | $-3.888 + 26.65i$ |
| 4 |  | $-3.456 + 9.576i$ | $-5.441 + 17.15i$ | $-4.744 + 22.68i$ | $-4.102 + 29.49i$ |
| $l$ |  | Spherical | | | |
| 0 |  | $-2$ | $-3$ | $-4$ | $-11.58$ |
| 1 |  | $-3$ | $-7.412$ | $-5.859 + 6.408i$ | $-4.167 + 14.72i$ |
| 2 |  | $-4.449 + 5.062i$ | $-7.038 + 10.85i$ | $-4.535 + 17.63i$ | $-4.243 + 24.76i$ |
| 3 |  | $-4.194 + 7.891i$ | $-7.401 + 14.97i$ | $-5.006 + 20.38i$ | $-4.450 + 27.63i$ |
| 4 |  | $-4.066 + 10.38i$ | $-7.268 + 18.84i$ | $-5.660 + 23.07i$ | $-4.693 + 30.40i$ |

Inspecting Fig. 1, we observe the following properties of the spectra. The real parts of all the eigenvalues are negative, $\operatorname{Re}\sigma < 0$, implying that the corresponding eigenmodes are stable: their amplitudes decay with time as $t^{-|\operatorname{Re}\sigma|}$. The decay is monotonic for negative real eigenvalues, which are few, and oscillatory in general. In the latter case, the corresponding eigenmode amplitude at late time decays as $t^{-|\operatorname{Re}\sigma|}\cos\varphi(t)$, where the phase $\varphi(t) = \operatorname{Im}\sigma \ln t + \varphi_0$, $\varphi_0$ being a constant, which means that the frequency of its oscillations decreases with time as $d\varphi/dt = \operatorname{Im}\sigma / t$.

The lowest mode $l = 0$ or $m = 0$ corresponds to a purely radial perturbation of the flow: the spherical or cylindrical symmetry of the background flow is not violated but, say, the shock location deviates from that predicted by the theory. There is no transverse motion, hence $D_1(\xi) \equiv 0$ for $j = 0$, cf. (27). The perturbation functions and the dispersion equation in this case become particularly simple for the spherical geometry, when the functions (24) are reduced to elementary functions:

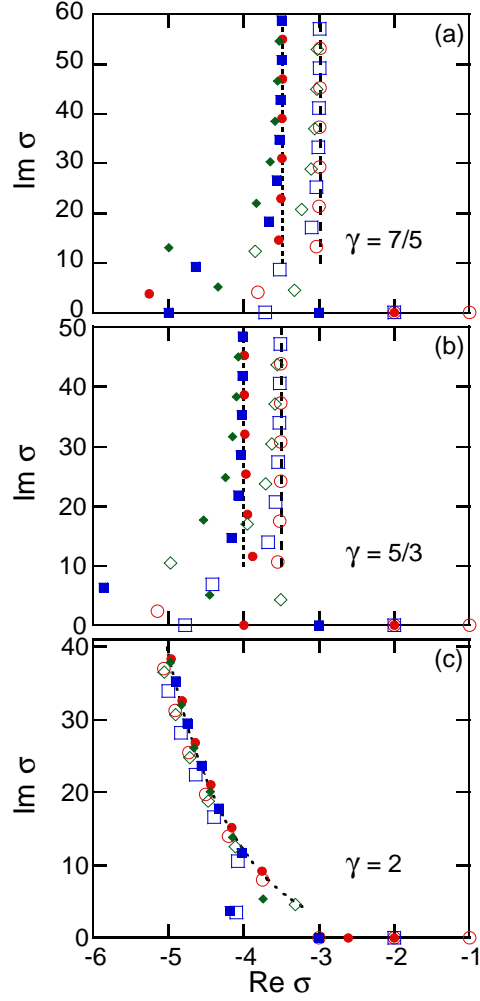

Figure 1. Spectra of the eigenvalues $\sigma$ shown in the non-negative $\operatorname{Im}\sigma \geq 0$ half of the complex plane for $\gamma = 7/5$ (a), $\gamma = 5/3$ (b) and $\gamma = 2$ (c). The eigenvalues corresponding to mode numbers $l, m = 0$, 1 and 2 are shown as circles, boxes, and diamonds, respectively. Filled and empty symbols correspond to the spherical and cylindrical geometry, respectively. Dashed and dotted vertical lines in (a) and (b) correspond to the asymptotic values (33) of $\operatorname{Re}\sigma$ with $\nu = 2$ and 3, respectively. The dotted line in (c) shows the asymptotic line (37).

$$f_+(\xi) = \frac{(1+M_2\xi)^{\sigma+1} - (1-M_2\xi)^{\sigma+1}}{2(\sigma+1)M_2\xi}, \tag{30}$$

and $f_-$ is obtained by replacing $\sigma+1$ with $\sigma+2$ in the right-hand side of Eq. (30). The dispersion relation (29) for the spherical geometry is reduced to

$$\left[ (2M_2+1)M_2^2\sigma^2 + (8M_2^2+M_2-2)M_2\sigma + 6M_2^3 - 4M_2^2 - 2M_2 + 2 \right](1+M_2)^{\sigma+1}$$
$$+ \left[ (2M_2-1)M_2^2\sigma^2 + (8M_2^2-M_2-2)M_2\sigma + 6M_2^3 + 4M_2^2 - 2M_2 - 2 \right](1-M_2)^{\sigma+1} = 0, \quad (31)$$

where we have substituted $\gamma = 1/(1-2M_2^2)$. Note that the value of $\sigma = -1$, at which the left-hand side of (31) vanishes, is not the root of the dispersion equation (29) because derivation of (31) involves multiplying its left-hand side by $\sigma + 1$.

For radial perturbations $l=0$ or $m=0$ there always exist two real eigenvalues. One of them, $\sigma = -\nu$, corresponds to a radial sonic wave because the vortical perturbations are excluded by the symmetry, and the entropic perturbation amplitude represented by the second term in the right-hand side of (26) vanishes identically. For this eigenmode, all the eigenfunctions are nonsingular and expressed via elementary functions:

$$P_1 = -(\nu-1)\left(\frac{\gamma+1}{2\gamma}\right)^{\frac{\nu-1}{2}}(1-M_2^2\xi^2)^{-\frac{\nu+1}{2}}, \quad G_1 = \frac{1}{\gamma}P_1,$$
$$V_1 = \frac{1}{\gamma}\xi P_1, \quad D_1 = 0. \quad (32)$$

The other real eigenvalue $\sigma = -\nu+1$ corresponds to an eigenmode involving an entropic wave. There could be more real eigenvalues, depending on the value of $\gamma$. For example, for spherical geometry one more real eigenvalue exists for $\gamma > 2$ and two more for $1.5565 < \gamma < 2$, see the Table.

Similarly, for the perturbations with $l=1$ or $m=1$ there always exists one real eigenvalue $\sigma = -\nu$ corresponding to a sonic eigenmode for which the entropic perturbation vanishes, cf. (26). But since this mode is not radial, the vortical contribution represented by the second term in the right-hand side of (27) is finite. For spherical geometry there can also be another negative real eigenvalue, which exists only for $\gamma < 2$. It tends to $\sigma = -4$ in the limit of high compressibility $\gamma \to 1$, which is easy to demonstrate by expanding the left-hand side of (29) in powers of small $\gamma - 1$. For an arbitrary $\gamma$ between 1 and 2 this eigenvalue is less than $-4$; e.g., for $\gamma = 7/5$ it equals exactly $-5$. For the cylindrical geometry the second real eigenvalue, which only exists for $\gamma < 2$, tends to $-3$ in the high-compressibility limit $\gamma \to 1$, and it is less than $-3$ for any $\gamma$ between 1 and 2.

Figure 1 demonstrates an infinite set of complex eigenvalues for any $l$, $m$, and $\gamma$. For the higher-order modes their real parts $\operatorname{Re}\sigma$ tend to a constant value, unless $\gamma = 2$ (see below), whereas their imaginary parts $\operatorname{Im}\sigma$ tend to infinity, asymptotically forming an arithmetic progression. We have calculated the limiting values of $\operatorname{Re}\sigma$ and the difference characteristic of this arithmetic progression, see Section D of Ref. 38:

$$\operatorname{Re}\sigma_n \cong -\frac{\nu - 1}{2} + \frac{\ln(|\mathscr{R}_s|)}{\ln \mathscr{D}_s} \cong \begin{cases} -\dfrac{\nu + 1}{2}, & \gamma \gg 1; \\ -\dfrac{\nu + 3}{2}, & \gamma - 1 \ll 1, \end{cases} \quad (33)$$

$$\operatorname{Im}\sigma_{n+1} - \operatorname{Im}\sigma_n \cong \frac{2\pi}{\ln \mathscr{D}_s} \cong \begin{cases} \dfrac{2\pi}{\ln(3 + 2\sqrt{2})}, & \gamma \gg 1; \\ \pi\sqrt{\dfrac{2}{\gamma - 1}}, & \gamma - 1 \ll 1. \end{cases} \quad (34)$$

Here the radial eigenmode number $n$ is a large positive integer, $n \gg 1$. The reflection coefficient for a planar sonic wave normally incident upon a shock front from downstream[18] is denoted by

$$\mathscr{R}_s = \frac{\delta p^{(s)}}{\delta p^{(0)}} = -\frac{1 - 2M_2}{1 + 2M_2} = -\frac{\sqrt{\gamma} - \sqrt{2(\gamma - 1)}}{\sqrt{\gamma} + \sqrt{2(\gamma - 1)}}, \quad -1 < \mathscr{R}_s < 3 - 2\sqrt{2} = 0.1716, \quad (35)$$

where $\delta p^{(0)}$ and $\delta p^{(s)}$ stand for pressure perturbation amplitudes in the incident and reflected sonic wave, respectively. [Eq. (35) corrects the typo on p. 342 of the English edition of Ref. 18.] The Doppler frequency reduction in the reflected sonic wave is given by the factor

$$\mathscr{D}_s = \frac{1+M_2}{1-M_2} = \frac{\sqrt{2\gamma}+\sqrt{\gamma-1}}{\sqrt{2\gamma}-\sqrt{\gamma-1}}, \quad 1 < \mathscr{D}_s < 3+2\sqrt{2} = 5.828. \qquad (36)$$

There is a simple heuristic derivation of the asymptotic equations (33)-(34). Let us assume that a normally incident spherical or cylindrical sonic wave reaches the shock front from downstream and is reflected back to the center or axis of symmetry at $t = t_0$, when the shock radius is $r_0 = M_2 c_s t_0$. Having been reflected from the center or axis, the sonic wave arrives to the shock front again at $t = t_1 = \mathscr{D}_s t_0$, when the shock radius is $r_1 = \mathscr{D}_s r_0$. There are two factors contributing to the attenuation of the sonic wave during the round trip: the reflection from the shock front and the divergence of the sonic wave front. The former is given by a factor $\mathscr{R}_s$ (35), the latter – by a factor of $(r_1/r_0)^{-\frac{1}{2}(\nu-1)} = \mathscr{D}_s^{-\frac{1}{2}(\nu-1)}$, see Ref. 41, Section 7.4. The total attenuation of the sonic wave therefore equals $\mathscr{R}_s \mathscr{D}_s^{-\frac{1}{2}(\nu-1)}$. On the other hand, it must also equal $(t_1/t_0)^\sigma = \mathscr{D}_s^\sigma$ to agree with the power-law time dependence assumed in (12)-(15). Calculating the logarithm of both sides of the equation $\mathscr{D}_s^\sigma = \mathscr{R}_s \mathscr{D}_s^{-\frac{1}{2}(\nu-1)}$, we arrive to (33)-(34). These formulas are valid asymptotically, in the limit of large radial mode number, because the planar formula (35) for the reflection coefficient and the above formula[41] for wave attenuation at large distances are only valid when the sonic wavelength is much less than the radius of the shock front, the condition ensured by $n \gg 1$. The above derivation explains why the asymptotic values (33), (34) depend on the geometry via the parameter $\nu$ determining the attenuation of diverging sonic waves, but do not depend on the mode number $l$, $m$, or mode wavelength, because neither the reflection coefficient $\mathscr{R}_s$ nor the Doppler shift factor $\mathscr{D}_s$ is wavelength-dependent.

For $\gamma = 2$ the limiting value (33) diverges because the reflection coefficient (35) vanishes, and the asymptotic behavior of the eigenvalues is different: their real parts tend to negative infinity for large radial eigenmode number. The asymptotic formula (34) is still valid. For spherical geometry and $l = 0$ the eigenvalues on the complex plane asymptotically approach the line defined parametrically by

$$\text{Re}\,\sigma_n = -\frac{1}{\ln 3}\ln\left(\frac{4\pi n}{\ln 3}\right) - 1, \quad \text{Im}\,\sigma_n = \frac{\pi}{\ln 3}\left(2n - \frac{1}{2}\right), \tag{37}$$

where $n \gg 1$ is now regarded as a continuous variable.[38] Figure 1(a), (b) and Tables I, II of Ref. 38 demonstrate that Eqs. (33), (34) are valid for both spherical and cylindrical geometries and for all mode numbers. Equation (37) is seen in Fig. 1(c) to be a good asymptotic approximation for all mode numbers in spherical geometry; the corresponding curve for cylindrical geometry exhibits the same logarithmic dependence of $\text{Re}\,\sigma_n$ on the mode number at $n \to \infty$.

A structure of the spectrum similar to that of Figs. 1(a), (b) – i. e., an infinite number of discrete eigenvalues $\sigma_n$, a constant negative real part of $\sigma_n$ and its imaginary parts forming an arithmetic progression, – has been found for the interaction of a planar shock wave with a rippled interface that triggers the development of the Richtmyer-Meshkov instability (RMI), see Ref. 42. Substituting $\nu = 1$ into (33), we essentially reproduce Eq. (71) of Ref. 42, whereas (34) is consistent with Eq. (72) of Ref. 42. Most studies of the small-amplitude RMI[42-44] concentrate on the evolution of the single unstable eigenmode but there is also an infinite set of stable sonic eigenmodes which exhibit an oscillatory decay as complex powers of time.

In Fig. 2 we plot some of the eigenfunctions representing the profiles of pressure, density, and radial velocity perturbations for spherical geometry, $\gamma = 5/3$, $l = 1$ for the first five radial eigenmodes. The radial mode numbers labeling the lines in Fig. 2 are the same as in the Table.

Since the expressions (26)-(28) contain the term $\xi^\sigma$ that diverges at the center (the physical meaning of this divergence is discussed later), $\xi = 0$ for all our stable eigenmodes, the profiles of $G_1$ and $V_{r1}$ are shown multiplied by $\xi^{-\sigma}$, which makes the resulting functions regular in the whole interval $0 \leq \xi \leq 1$. For the two lowest radial eigenmode numbers, $n = 1$ and 2, the corresponding eigenvalues are real, and so are the eigenfunctions. For the other three modes the eigenvalues and the eigenfunctions are complex, and we show in Fig. 2 only the real parts of these eigenfunctions.

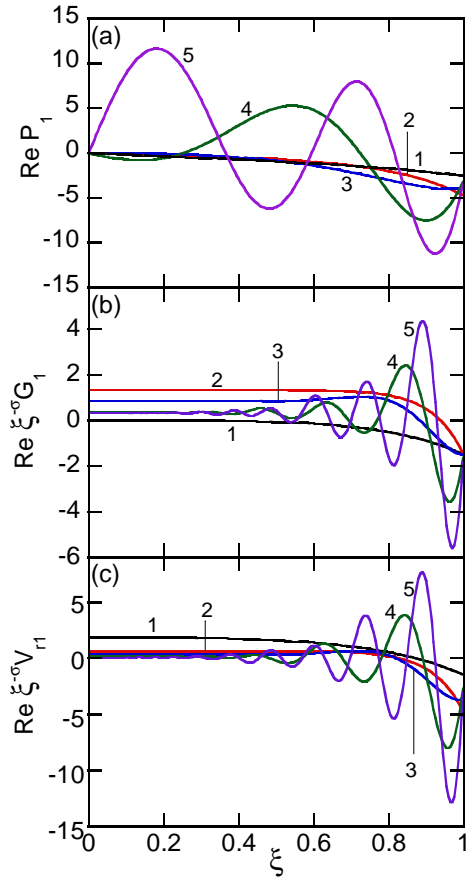

Figure 2. Eigenfunction profiles of the normalized perturbations of pressure (a), density (b) and radial velocity (c) for spherical geometry, angular mode number $l = 1$ and $\gamma = 5/3$. Only the real parts of the eigenfunctions are shown. The lines are labeled by the eigenmode numbers, same as in the Table.

Inspecting Fig. 2(a), it is easy to understand why the number of the radial eigenmodes has to be infinite. The sonic contribution of the $n^{\text{th}}$ radial perturbation mode, Fig. 2(a), is a standing sonic wave, whose real and imaginary parts both have exactly $n-2$ (for $n \geq 3$) nodes between the center and the shock front. We obviously need all these standing waves to decompose an arbitrary pressure perturbation imposed at some finite initial moment $t = t_0 > 0$ into an infinite sum of the eigenfunctions of our perturbation problem. It is also clear why these standing sonic waves decay with time. They reverberate in a space bounded by the expanding shock wave, reflecting from this expanding surface from behind. The reflected sonic waves are thereby Doppler-shifted to lower frequencies, that is, to lower energies, by a large factor of $\mathscr{D}_s$, see (36). Moreover, the reflection coefficient $\mathscr{R}_s$ (35) is quite low if the adiabatic index $\gamma$ is not too close to unity, and for oblique incidence the reflection coefficient is even lower. The above two factors ensure the rapid decay of the sonic wave amplitudes for all eigenmodes. Since the pre-shock gas is unperturbed, the perturbations come to the shock front only from behind as sonic waves, hence all the perturbation amplitudes near the shock front exhibit a rapid decay as the shock front expands.

In addition to the sonic contributions, which are finite everywhere, the profiles of density and velocity include the contributions of the entropy and vorticity perturbations, respectively. These perturbations do not propagate through the shocked gas, they are localized in the fluid particles. Their contributions are proportional to $\xi^\sigma$ and therefore diverge as $\xi^{-|\text{Re}\,\sigma_n|}$ at $\xi \to 0$ while oscillating with a frequency that increases near $\xi = 0$ as $\text{Im}\,\sigma_n / \xi$.

Notwithstanding this divergence, all the eigenfunctions (26)-(28) are physically meaningful. The above divergence is a direct consequence of the separation of variables $\xi$ and $t$ assumed in our perturbation equations. The entropy and vorticity contributions to the density and velocity profiles, which identically satisfy the perturbation equations (16)-(19), are arbitrary time-independent functions $F(r)$ of the radius $r$. But since its argument is $r = v_s t \xi$, and the separation of variables imposes the same time dependence $t^\sigma$ on all the perturbation functions, the function $F(r)$ for each eigenmode has to be proportional to the same power of its argument, that is, to $(t\xi)^\sigma$.

This divergence is due to the singular nature of our perturbation problem at the time origin, $t = 0^+$, when the radius of the shock front is zero. The small-amplitude assumption, on which the theory is based, requires the shock displacement amplitude to be much smaller than $R_s$ but this requirement clearly cannot be satisfied at the initial instant. We have essentially the same situation in the case of RMI.[42] The difference is that in the latter case, we have one unstable eigenmode that is regular at $t \to 0^+$. There is a single radial eigenmode that describes the instability development, and all the studies of the classical RMI focus on it;[42-44] other modes, which are diverging at $t \to 0^+$ and thereby singular at $\xi \to 0$, rapidly decay with time and do not affect the numerical results or analytical late-time asymptotics.

Stability of the Noh solution implies that such perturbation eigenmodes are the only ones that exist. Our theory is therefore only applicable starting from some finite instant of time, which we can identify with our time unit introduced in (11): $t \geq t_0 > 0$. Then, formulating the initial conditions that correspond to a particular $(l, m, n)$ eigenmode at $t = t_0$, we can take the exact pressure perturbation profile given by (25) as the initial condition for the pressure. All the other perturbation functions are singular at $\xi \to 0$ unless we deal with the radial eigenmode that corresponds to $l = 0$ or $m = 0$ and $\sigma = -\nu$, see (32). In the other perturbation functions, we replace the diverging terms $\xi^\sigma$ with arbitrary functions of $\xi$ defined on the interval $0 \leq \xi \leq 1$. These functions have to be regular at $\xi = 0$ and smoothly join the exact eigenmode profiles at later instant of time – the two requirements that leave a high degree of arbitrariness in their choice. The profiles constructed this way satisfying all the boundary conditions at the shock front at $t = t_0$. At $t > t_0$ difference in the density/entropy and vorticity perturbations would not affect the continuously added shocked gas because these are non-propagating perturbations localized at the fluid particles and do not interact with similar perturbations in the particles shocked later. As the time increases, the radius where the numerical solution deviates from our exact solution decreases, when compared to the shock radius, as $t_0/t$. For $t \gg t_0$ the profiles of the fluid variables reproduce our eigenfunctions in most of the shocked volume with the exception of the immediate vicinity of the center or the axis of symmetry, where the solution constructed this way is regularized, i. e. non-singular. Such solution could be used for a 2D or 3D code verification, which is itself a non-trivial problem, as demonstrated in the next Section.

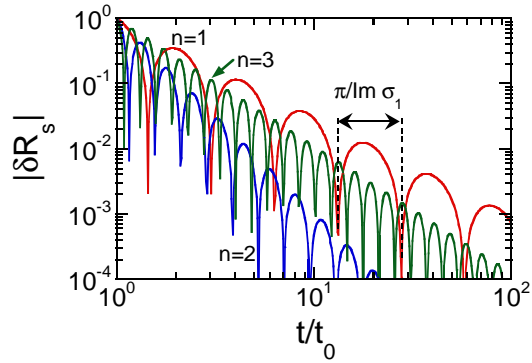

Figure 3. Relative shock front displacement amplitude vs. normalized time for cylindrical geometry, $\gamma = 5/3$, $m = 2$ and the three lowest eigenvalues listed in the Table.

Figure 3 illustrates the time history of the normalized relative shock displacement amplitude $\delta R_s = \delta r_s(t)/R_s = \mathrm{Re}(t^\sigma)$ for the particular case of cylindrical geometry, $m = 2$ azimuthal mode number, $\gamma = 5/3$ and the three lowest complex-conjugate eigenvalues with radial mode numbers $n = 1, 2, 3$ in the Table. All the mode amplitudes are normalized to unity at $t = t_0$. Plotting the absolute value of $\delta R_s$ makes it possible to use the log-log scale, illustrating the rapid decay of these perturbations ensured by the large negative real parts of $\sigma_n$. Frequency of the oscillations is seen to increase with $n$ as a result of the increased imaginary part of $\sigma_n$. The frequency, which for a given $n$ appears as constant on the log time scale of Fig. 3, decreases with time as $1/t$. For the given mode number $n$, the increase of $\ln t$ corresponding to sequential zeroes of $\mathrm{Re}(t^\sigma)$ equals $\Delta \ln t = \pi / \mathrm{Im}\,\sigma_n$. For the lowest-order mode $n = 1$ in Fig. 3 we have: $\pi / \mathrm{Im}\,\sigma_1 = 0.7385$, and the nodes of $\mathrm{Re}(t^\sigma)$ form a geometric progression: between the two subsequent nodes the time increases by a constant factor equal to $e^{0.7385} = 2.093$ for $n = 1$, as seen in Fig. 3. The same applies to all the other eigenmodes oscillating at higher frequencies.

## III. NUMERICAL SOLUTION OF THE NOH PROBLEM

### A. Simulations on a Cartesian grid

As a first step towards using the above analytic solution for hydro code verification, we have simulated the Noh expanding-shock flow starting from either $t = 0$ or a finite moment of time, $t = t_0$. The simulations were done for cylindrical and spherical geometry, in two and three dimensions, respectively, on a uniform Cartesian grid. No 2D or 3D perturbations have been initially imposed. The Cartesian grid itself provides a constant input of perturbations through the expanding shock front due to the non-alignment between the theoretical shape of the shock front, spherical or cylindrical, and the cell boundaries oriented along the Cartesian axes. Here we are examining whether a particular code (Athena) running on a uniform Cartesian mesh is a viable candidate for the use of this analytic solution as a verification tool. We know that attempting to use a Cartesian mesh to simulate a problem for which the unperturbed solution has cylindrical or spherical symmetry will result in numerically induced noise, even with no initially imposed perturbations. The first question before us then is whether this numerically induced noise will decay at a rate faster than the rate at which an initially imposed perturbation would decay physically. If so, then this code is a candidate for verification via the analytic solution we are introducing here. If not, then this code, when run on a Cartesian mesh, is probably not a good candidate for examining decaying perturbations of the type addressed here.

The simulations were done using a version of the Athena code.[45, 46] This Eulerian code is based on a high-order Godunov finite-volume discretization. The fluid variables are carried at the cell centers and the code conserves mass, momentum, and total energy. This code had originally been developed for astrophysical studies and is actively supported by a growing community. In fact, our choice of a Cartesian mesh rather than a cylindrical one was dictated in part by the confidence that we had developed in the Cartesian version of Athena during our previous work on the MHD version of the Noh problem,[32] and in part out of our curiosity about whether such a high-performance code could properly resolve decaying perturbations even on a non-aligned mesh. Our original intent had been to subsequently exercise the newly-added cylindrical mesh capabilities of Athena for this test, but we encountered some numerical difficulties with that effort which we have not yet been able to resolve. We will attempt that exercise, or a similar one, at a later date.

We have chosen $\gamma = 5/3$, $v_0 = 1$, and $\rho_0 = 1$, so that the expanding shock velocity (6) is $v_s = 1/3$. For cylindrical and spherical geometry, $\nu = 2$ and 3, we have the post-shock values $\rho_s = 16$, $p_s = 16/3$, and $\rho_s = 64$, $p_s = 64/3$, respectively, see Eqs. (7) and (8). In the simulations the latter values are approximate rather than exact because the pre-shock pressure is not exactly zero, as in (4). We have taken $p_0 = 10^{-6}$, which implies a deviation from the exact solution only in the 5$^{th}$ and higher digits. The computational box is symmetric, $[-L, L]^2$ and $[-L, L]^3$ for 2D cylindrical and 3D spherical geometry, respectively, with $L = 3.5$. The resolution of a uniform square Cartesian grid for cylindrical geometry was $N_c^2$, with the number of cells per axis $N_c$ varied from 256 to 4096, and for spherical geometry it was $N_c^3$, with $N_c$ varied from 128 to 768. The Courant number was kept fixed at 0.2 in both cases. When the simulation run started from a finite moment $t = t_0$, the initial conditions for all fluid variables were determined by the exact 1D self-similar Noh solution given by the formulas (5) and (9) by point-wise evaluation at grid cell centers. The Athena code was configured to use the van Leer integrator, third-order reconstruction in primitive variables, and the Roe Riemann solver with H-correction. The version of Athena provided by Jim Stone[47] allows the H-correction to be used with the van Leer integrator, whereas the latest public release, version 4.2, does not.

Figures 4 to 6 present the simulated maps of pressure, density and radial velocity for the case of 2D cylindrical geometry initiated at the finite instant of time, $t_0 = 2$, when $R_s = 2/3$, and simulated with $N_c = 1024$. The color ranges in all maps are adjusted to emphasize the post-shock solutions. The pressure maps of Fig. 4 show sonic perturbations, which are particularly noticeable for modes $m = 4$ and 8. As the shock front expands, the perturbed pressure field becomes smoother: the amplitudes of these perturbation modes decrease, although they remain observable in the vicinity of the shock front. Figure 5 shows the simulated density maps for the same conditions. Note the difference between Figs. 4 and 5. The density map includes the contributions from the sonic waves and entropy perturbations. The former perturbations decay with time, whereas the latter ones stay constant in the shocked fluid particles. We clearly recognize in Fig. 5 the perturbations of both kinds. The entropy perturbations contribute the recognizable steady patterns with $m = 4$ and 8 symmetry. They are particularly large near the radial position where the numerical solution was initialized. This is a signature of the initial conditions imprinted on the grid before the numerical solution started approaching the theoretical solution. The sonic perturbations are responsible for the oscillatory wave patterns, which are mostly visible near the shock front. Similar perturbation profiles are recognizable in the maps of radial velocity shown in Fig. 6 for the same conditions. Here the steady patterns are due to the vorticity perturbations that stay constant in the fluid particles.

The general decay of perturbations in the shocked gas with time is best characterized by the pressure perturbation amplitude. For any eigenfunction, the latter decays with time as $t^{-|\text{Re}\,\sigma|}$, the negative power being 3 or greater, see the Table and Fig. 3. To evaluate the average amplitude of the relative root-mean-square (rms) pressure perturbation, we use the following metric:

$$\delta p_{rms} = \sqrt{\frac{1}{N_s} \sum_{i=1}^{N_s} \left( \frac{p_i}{p_s} - 1 \right)^2}, \qquad (38)$$

where $p_s = 16/3 = 5.333$ is the theoretical post-shock pressure given by (8), and $p_i$ is the pressure in the $i^{\text{th}}$ grid cell. The summation in (38) is done over all the $N_s$ grid cells fully contained between $r = 0$ and $r = R_s - 2\Delta x$, where $r = R_s$ is the theoretical shock front position and $\Delta x = 2L/N_c$ is the grid spacing. Figure 7(a) shows the perturbation metric (38) vs. time for the simulation runs done starting from $t = 0$ in cylindrical geometry with the highest grid resolution $N_c = 4096$. The value of $\delta p_{rms}$ is presented for

$$t > \frac{2L}{v_s N_c}\left(2+\sqrt{\nu}\right), \tag{39}$$

when at least one grid cell per quadrant is fully contained inside the radius $r = R_s - 2\Delta x$, so that a meaningful evaluation of (38) can be made. For lower resolution this instant will be accordingly later. We have checked that the value of $\delta p_{rms}$ is independent of resolution and fully determined by the number of the grid cells fully contained behind the expanding shock front, that is, by the ratio $R_s / \Delta x$. The deviation of the numerical perturbation field from the 1D theory is not due to a random noise, like that generated by round-off errors, but rather it is a robust feature of the finite-difference 2D approximation of the original Noh problem. On the log-log scale of Fig. 7(a), the curve $\delta p_{rms}(t)$ asymptotically approaches a straight line, indicating a power-law time dependence. The asymptotic power index estimated for these simulation results is $-1/2$, which is illustrated by the straight dotted line: $\delta p_{rms}(t) \propto t^{-1/2}$. The decay illustrated by Fig. 7(a) is thereby slower than $t^{-1}$, the power law characteristic of the radial eigenmode $m = 0$ corresponding the lowest value of $\sigma = -1$ for cylindrical geometry, cf. Fig. 1.

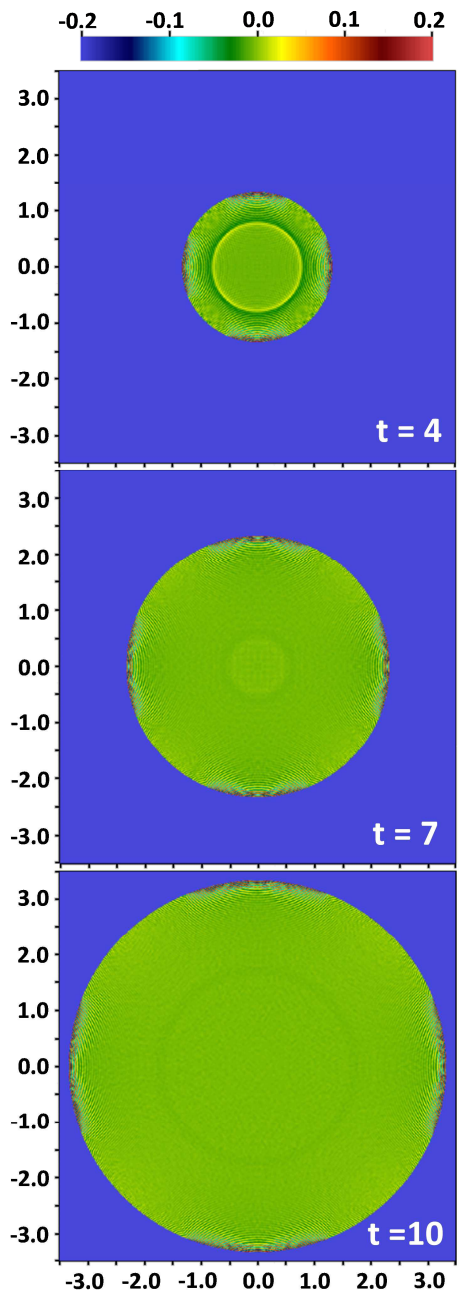

Figure 4. Simulated postshock pressure maps for cylindrical geometry, $\gamma = 5/3$ and the normalized time increased from $t = 4$ to $t = 10$.

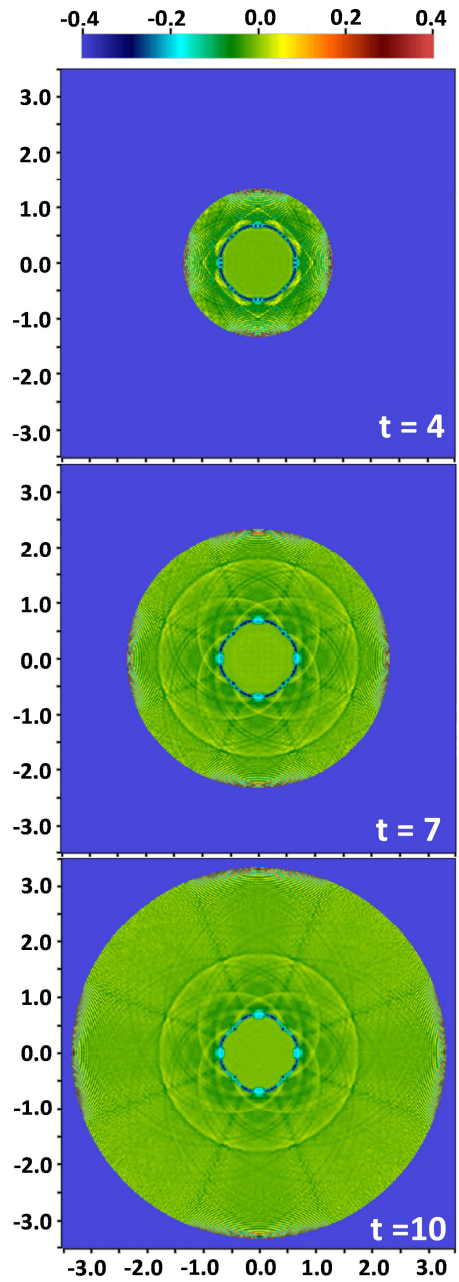

Figure 5. Simulated postshock density maps for cylindrical geometry, $\gamma = 5/3$ and the normalized time increased from $t = 4$ to $t = 10$.

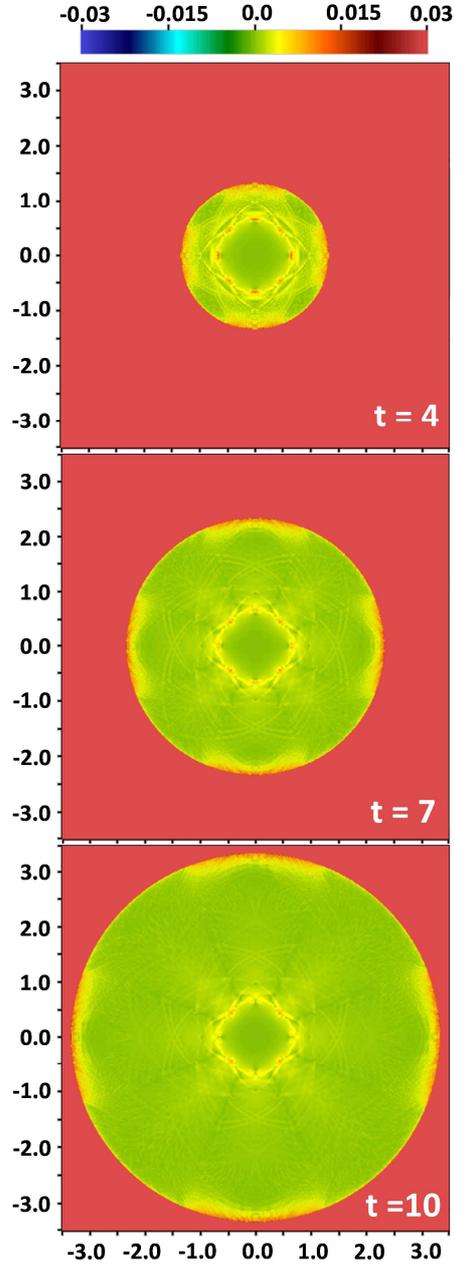

Figure 6. Simulated postshock maps of radial velocity for cylindrical geometry, $\gamma = 5/3$ and the normalized time increased from $t = 4$ to $t = 10$.

A similar behavior has been detected when the same Athena code was used for 3D simulations of the spherical Noh problem. Figure 8 shows the simulated density map on the $(x, y)$ plane at $t = 10$ obtained in a simulation done for $\gamma = 5/3$, started at $t_0 = 2$ (the same conditions as in the 2D cylindrical simulations illustrated by Figs. 4-6), but performed in 3D spherical geometry with a lower resolution, $N_c = 128$. Qualitatively it is not different from the similar map of Fig 5. The wave patterns due to the sonic perturbations are more visible here due to the lower resolution. The steady structures due to entropy perturbations are also prominent, more so near the area of the shock initiation.

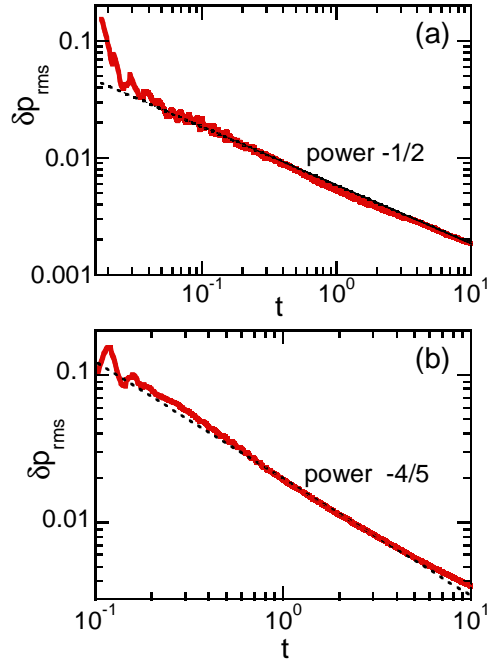

Figure 7. Relative rms pressure perturbation amplitude $\delta p_{rms}$ vs. time for simulations run on Cartesian grid for $\gamma = 5/3$. (a) 2D cylindrical geometry, resolution $4096^2$; (b) 3D spherical geometry, resolution $768^3$. Straight dotted lines in (a) and (b) correspond to the slope $\propto t^{-1/2}$ and $\propto t^{-4/5}$, respectively.

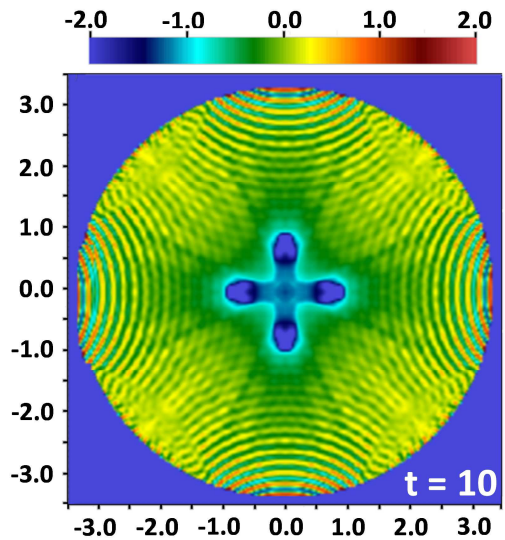

Figure 8. Simulated postshock density map for the 3D spherical geometry, $\gamma = 5/3$, resolution $128^3$ and the normalized time $t = 10$.

Figure 7(b) shows the perturbation metric (38) with the appropriate value of $p_s = 64/3 = 21.333$ vs. time for the simulation run for $\gamma = 5/3$ in 3D spherical geometry with the highest resolution $N_c = 768$. The results are shown starting for the time given by (38) with $\nu = 3$ when at least one grid cell per octant is fully contained inside the radius $r = R_s - 2\Delta x$, so that a meaningful evaluation of (37) can be made. Here again, we have checked that the perturbation metric (37) is a function of the number of the grid cells fully contained behind the expanding shock front, and it is independent of the resolution. On the log-log scale of Fig. 7(b), the curve $\delta p_{rms}(t)$ asymptotically approaches a straight line. However, the power index here is not the same as for the cylindrical geometry, Fig. 7(a). Rather, it is approximately $-4/5$, which is illustrated by the straight dotted line $\delta p_{rms}(t) \propto t^{-4/5}$. Here again, the decay illustrated by Fig. 7(b) is slower than $t^{-2}$, the power law characteristic of the radial eigenmode $l = 0$ corresponding the lowest value of $\sigma = -2$ for cylindrical geometry, cf. Fig. 1.

## B. Simulations on a curvilinear grid in spherical coordinates

Numerical solution of the Noh problem on a curvilinear grid in spherical geometry has been conducted in 2D, assuming azimuthal symmetry. The symmetry enables us to calculate in a two-dimensional computational grid and save computational resources. The curvilinear coordinates of the axes were in radial and polar directions. The cell size was uneven in the radial direction, so that the ratio between the cell size and the radius of its center was constant. The radius of the $i^{th}$ node was expressed as

$$r_i = r_{in} \exp\left( \frac{i}{i_{max} - 1} \ln \frac{r_{out}}{r_{in}} \right) \tag{40}$$

where $r_{in}$ and $r_{out}$ are, respectively, the inner and outer boundaries of the computation domain, $i_{max}$ is the number of grid nodes in the radial direction. We have chosen $r_{in} = 0.001$ and $r_{out} = 1$, and the number of grid nodes was up to 1233 and 513 in the radial and polar directions, respectively. A source term related to the pressure force acting on different radial positions was introduced in the radial momentum equation.

Convective terms were evaluated using the SLAU2 scheme[48] that provides robust modeling of strong shock wave propagation and has been used in our previous study.[37] However, preliminary calculations had shown that the original SLAU2 scheme produced spurious rarefaction waves emitted from the shock front to the center. We found that this effect can be suppressed by increasing the numerical viscosities, especially in the proximity of the shock wave. The numerical mass flow $\dot{m}$ and pressure flux $\tilde{p}$ adopted in the present calculations were expressed by following equations.

$$\dot{m} = \rho_L \frac{v_{n,L} + |v_{n,L}|}{2} + \rho_R \frac{v_{n,R} - |v_{n,R}|}{2} + s\left[(1-g)\frac{\rho_L \rho_R}{\rho_L + \rho_R}\left(|v_{n,R}| - |v_{n,L}|\right) + \frac{\chi}{2c_{1/2}}(p_L - p_R)\right], \quad (41)$$

$$\tilde{p} = \max\left\{\frac{p_L + p_R}{2} + \frac{f_p^+ - f_p^-}{2}(p_L - p_R) + \left[\sqrt{\frac{|\mathbf{v}_L|^2 + |\mathbf{v}_R|^2}{2}}\left(f_p^+ + f_p^- - 1\right) + 0.2(v_{n,L} - v_{n,R})\right]\right.$$
$$\left. \times \frac{\rho_L + \rho_R}{2}\min(c_L, c_R), \quad \min(p_L, p_R)\right\}, \quad (42)$$

$$s = 1 - \min\left\{1, \quad \max\left[0, \quad M_{n,L} - M_{n,R} + \min\left(\left|\frac{\rho_L}{\rho_R} - 1\right|, \left|\frac{\rho_R}{\rho_L} - 1\right|\right)\right]\right\}. \quad (43)$$

Here $\mathbf{v}$, $v_n$, $\rho$, $p$ and $c$ are, respectively, the velocity vector, the velocity component normal to the cell boundary, density, pressure and speed of sound; $M_n$ is the Mach number evaluated by the velocity normal to the boundary. The 2nd order MUSCL interpolation has been applied to the density, velocities and temperature. The other values, e.g. the pressure, are calculated from the interpolated values. Subscripts L and R indicated which side of the grid points was utilized for the MUSCL interpolation, and $g$, $f$ and $\chi$ are switching parameters depending on the Mach number defined in the original reference.[48] We have introduced the new parameter $s$ that detects the strong moving shock as seen in the Noh problem. The parameter $s$ controls the numerical diffusion in order to avoid numerical generation of spurious rarefaction waves. The three-points backward difference (second order in time) was adopted for the temporal terms. The time step utilized was $2 \times 10^{-5}$, for which the maximum CFL number was about 2.6.

Numerical calculations have been carried out for the background conditions similar to those of Section III.A: specific heat ratio $\gamma = 5/3$, initial velocity $v_0 = 1$, initial density $\rho_0 = 1$, and initial pressure $p_0 = 10^{-6}$. The initial shock radius was modulated via the Legendre polynomial of the order $l$:

$$r_s(t = t_0, \theta) = r_{s,init} + \delta r_s P_l(\cos\theta). \tag{44}$$

We set the initial shock position without perturbation $r_{s,init}$ to $0.1 \times r_{out}$, and the modulation amplitude of the shock position either zero or $\delta r_s = 0.05 \times r_{s,init}$. No initial perturbation other than shock position was applied in the present calculations.

The pressure perturbation metric (38) has been modified as the weighted sum

$$\delta p_{rms} = \sqrt{\frac{1}{\sum_i V_i} \sum_i V_i \left(\frac{p_i}{p_s} - 1\right)^2}, \tag{45}$$

where $V_i$ is the volume of the $i^{th}$ grid cell, because here these volumes are different. The summation over $i$ in (45) is done over all the grid cells fully contained behind the expanding shock front. Similarly, the relative rms displacement of the shock front is defined by the weighted sum

$$\delta R_{rms} = \sqrt{\frac{1}{\sum_j S_j} \sum_j S_j \left(\frac{r_{s,j}}{r_{s,av}} - 1\right)^2} \,, \quad (46)$$

where $S_j$ is the outer surface area of the $j^{th}$ grid cell through which the shock front passes at the given instant of time, $r_{s,j}$ is the shock radius corresponding to this cell, and $r_{s,av}$ is the average shock radius at the given instant of time.

Figure 9 shows the relative rms pressure perturbation amplitudes $\delta p_{rms}$ vs. time for the simulations run for the radially-symmetric case of $l = 0$ (a) (for this case we have chosen $\delta r_s = 0$ in (44)) and $l = 1$, 2, 3, and 4. The time is shown normalized with respect to $t_0 = r_{s,init} / v_s = 0.3$, where $v_s = 1/3$ is given by (6). We see that in all cases the value of $\delta p_{rms}$ does not decay as in Fig. 7; rather, it fluctuates between 0.1 and 0.01 with a constant frequency. These sonic perturbations being constantly added to the flow are apparently due to the new grid cells joining the shocked area. As the shock front propagates outward, the ratio of the grid cell size to the shock front radius does not decrease with time, which explains the difference with the results of Section III.A. Figure 9 demonstrates that the numerical noise is essentially the same for all values of $l$ and is thereby mainly due to the purely radial, $l = 0$ deviation of the numerical solution from the theory.

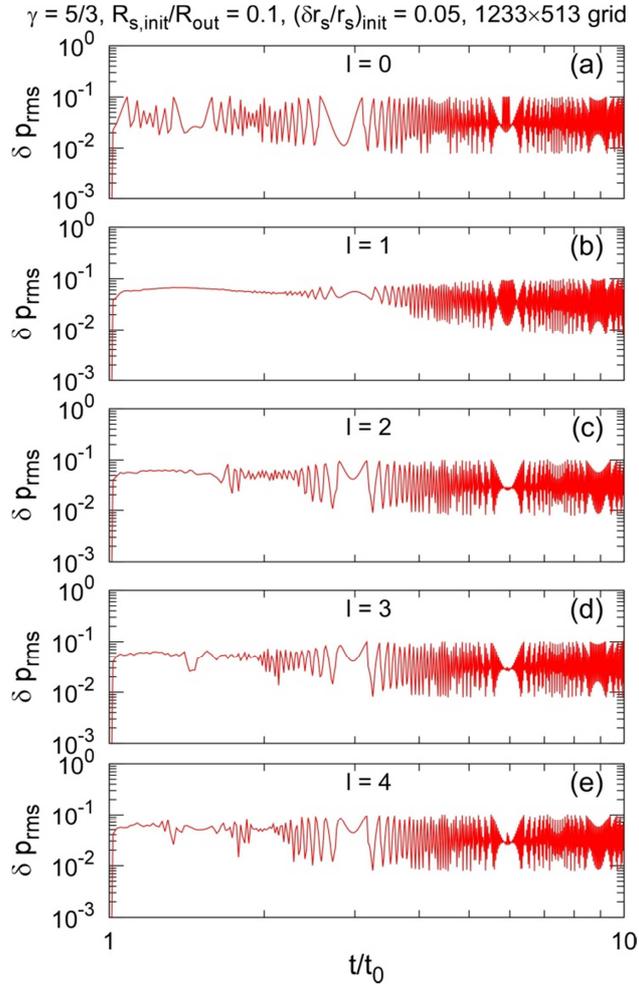

Figure 9. Relative rms pressure perturbation amplitude $\delta p_{rms}$ vs. time for simulations run on a 2D curvilinear grid in spherical coordinates for $\gamma = 5/3$. (a) $l = 0$, $\delta r_s = 0$ in (44); (b), (c), (d) and (e) correspond to $l = 1, 2, 3$, and $4$, respectively, and $\delta r_s / r_{s,init} = 0.05$.

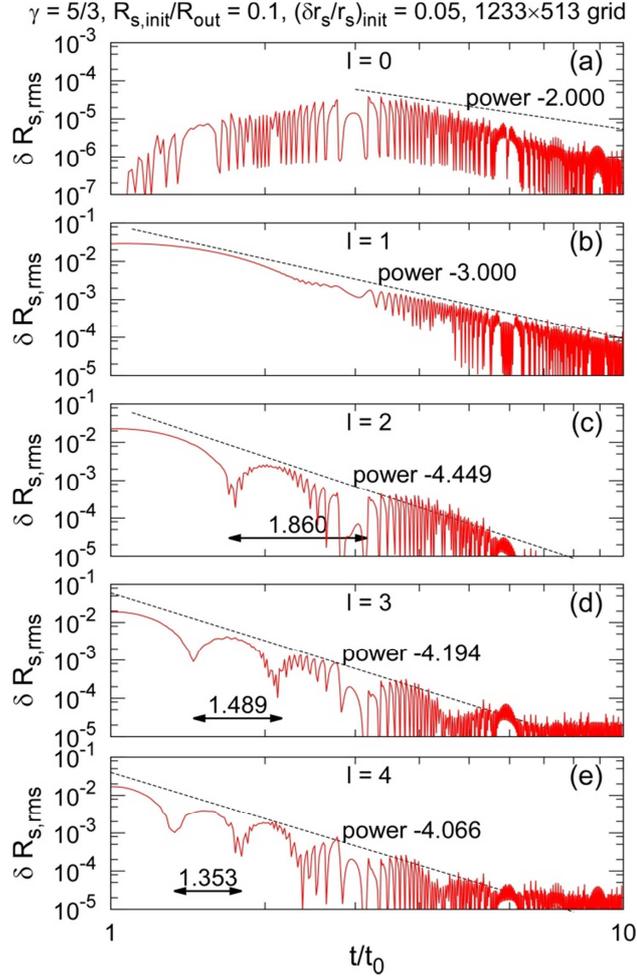

Figure 10. Relative rms shock displacement amplitude $\delta R_{s,rms}$ vs. time for simulations run on a 2D curvilinear grid in spherical coordinates for $\gamma = 5/3$. (a) $l = 0$, $\delta r_s = 0$ in (44); (b), (c), (d) and (e) correspond to $l = 1$, 2, 3, and 4, respectively, and $\delta r_s / r_{s,init} = 0.05$. Straight dotted lines correspond to the slope $\propto t^{\mathrm{Re}\,\sigma_1}$ where $\sigma_1$ is the lowest-order $n = 1$ eigenvalue for a given $l$ in the Table. The double arrows in (c), (d) and (e) indicate the relative time increase between the two sequential zeroes of the lowest-order mode, $\exp(\pi / \mathrm{Im}\,\sigma_1)$.

Figure 10 demonstrates, for the same conditions as Fig. 9, the simulated relative rms displacement of the shock front vs. time. According to the theory, the relative shock displacement amplitude depends on time as $t^\sigma$ for each radial mode. The power indices $\mathrm{Re}\,\sigma_1$ corresponding to the lowest-order modes corresponding to $n = 1$ in the Table for $l = 0$, 1, 2, 3, and 4 are, $-2$, $-3$, $-4.449$, $-4.194$, and $-4.066$, respectively. The corresponding slopes are illustrated in Fig. 10 by thin straight lines. The $n = 1$ eigenmodes for $l = 0$ and 1 are non-oscillatory. For $l = 2$, 3 and 4 the denominator of the geometric progression characteristic of the oscillations of the shock displacement amplitude is $\exp(\pi / \operatorname{Im} \sigma_1) = 1.860$, 1.489, and 1.353, respectively. These intervals are shown in Figs. 10(c), (d) and (e) as the time scale of the lowest-frequency oscillations, cf. Fig. 3.

Given that the radial $l = 0$ pressure perturbation amplitude does not decay with time, see Fig. 9, we cannot expect the simulated shock displacement amplitude to decay with time according to the theoretically predicted law $\propto t^{-2}$. The amplitude shown in Fig. 10(a) basically stays constant, which is consistent with Fig. 9(a). The magnitude of $\delta R_s$ in Fig. 10(a) is much smaller compared to the other parts of Fig. 10, where the simulated shock perturbations with the angular dependence given by $l = 1$, 2, 3 and 4 are demonstrated to decay with time, in a reasonable agreement with the above power laws predicted by the theory. The half-periods of the low-frequency oscillations in Figs. 10(c), (d) and (e) are seen to be consistent with the above values of $\exp(\pi / \operatorname{Im} \sigma_1)$. These results, together with those of Section III.A, provide another confirmation of the stability of the classic Noh solution. Moreover, they demonstrate the decay of shock front perturbation amplitudes for the modes with $l = 1$, 2, 3 and 4 in a general agreement with the predictions of the theory. This indicates a possibility of using our theory for more detailed code verification.

## IV. CONCLUSIONS

We have presented a stability analysis for the simplest case of stagnation via an expanding accretion shock wave represented by the classic Noh solution in spherical and cylindrical geometry. The stagnation has been demonstrated to be stable for any adiabatic index $\gamma$ and all mode numbers $l$ and $m$. All the perturbation eigenmodes decay with time as powers of $t$. The decay is oscillatory for most eigenmodes, the oscillation frequency decreasing with time as $1/t$. Our present results represent the necessary first step in a systematic stability study of the 1D flows that describe stagnation via accretion shock, which by definition are partly supported by the incident plasma flow, but not necessarily fully supported, as in the classic Noh case, which opens the possibility for an instability development. The most relevant for ICF/HEDP example of such accretion-shock flow is the reflected-shock phase of the Guderley solution.[34-36] Purely gasdynamic stability analysis, like that done here and in Ref. 37, is, in turn, a pre-requisite for a more complete analysis accounting for additional physics, including MHD effects,[32,33] radiative energy losses from the stagnated cylindrical precursor plasma,[12,13] or both radiative losses and gravity in astrophysical accretion-shock flows.[17] Such study is obviously necessary for the analysis of the laboratory-astrophysics experiments proposed in Ref. 16.

The physical stability of the classic Noh solution in two and three dimensions ensures the lasting value of this solution as a 1D verification test for hydrocodes. We have presented some examples of such numerical verification in two and three dimensions for cylindrical and spherical geometry. Our numerical solutions obtained on a Cartesian grid using the Athena code have been demonstrated to converge to the classic Noh solution as the value of the ratio $R_s/\Delta x$ increases with time, as should occur. The 2D or 3D numerical solution asymptotically approaches the 1D theoretical solution but deviates from it due to the non-alignment between the theoretical shape of the shock front, cylindrical or spherical, and the grid cell boundaries, which are parallel to the Cartesian axes. As the 2D or 3D sonic perturbations decay while reverberating inside the shocked gas, new perturbations are being added through the shock front, thereby determining the overall perturbation dynamics. Choosing the relative rms pressure perturbation amplitude $\delta p_{rms}$ as a metric of the difference between the numerical and theoretical solutions, we have found that $\delta p_{rms}$ decays with time as $\sim t^{-1/2}$ and $\sim t^{-4/5}$ for 2D cylindrical and 3D spherical geometry, respectively. The above power indexes are lower than those characteristic of the physical eigenmodes corresponding all the relevant mode numbers starting from $m = 0$ and $l = 0$. Therefore the "perturbation field" in our simulations is totally determined by the discreteness of the grid. In other words, successful code verification in 1D does not imply that the 2D or 3D perturbation field is also simulated accurately.

Our 2D simulations done on a curvilinear grid in spherical coordinates with the an improved version of the code used in Ref. 37 do not reproduce the decay of $\delta p_{rms}$ with time because the dimensions of the grid cells reached by the shock front do not become smaller compared to the shock front radius as it expands. Nevertheless, the simulated decay of the amplitudes and oscillation frequencies of the relative rms shock-front displacement for radial modes with $l = 1$, 2, 3 and 4 are found to be in a general agreement with the theory, which is encouraging.

The exact analytic expressions obtained here for the eigenmodes in principle open a possibility for using the exact solutions of the perturbed Noh problem for verification of hydrocodes. Our numerical examples demonstrates that this is a non-trivial task because in the Cartesian simulations it is hard, if not impossible, to resolve the rapidly decaying physical eigenmodes on top of the numerical deviation from the 1D solution that decays much slower. It would be interesting to find out if a different numerical scheme, e. g. one using cylindrical or spherical coordinates, can reduce sufficiently the effect of grid discreteness on the simulations of the Noh problem, so that the dynamics of the physical eigenmode could be resolved. This question remains open for future studies.

## ACKNOWLEDGMENTS

Work supported by the US Department of Energy/NNSA, and by the Japan Society for the Promotion of Science. The authors wish to thank Prof. Jim Stone of Princeton University for providing the version of Athena used for the numerical simulations presented in this paper. A. L. V. gratefully acknowledges the Institute of Laser Engineering, Osaka University, for their support and hospitality during his visit, when most of the theoretical results presented here have been obtained.

## REFERENCES

[1] J. Lindl, *Inertial Confinement Fusion: The Quest for Ignition and Energy Gain Using Indirect Drive* (AIP-Press, New York, 1998).

[2] S. Atzeni and J. Meyer-ter-Vehn, *The Physics of Inertial Fusion: Beam Plasma Interaction, Hydrodynamics, Hot Dense Matter* (Clarendon Press, Oxford, 2004).

[3] R. S. Craxton, K. S. Anderson, T. R. Boehly, V. N. Goncharov, D. R. Harding, J. P. Knauer, R. L. McCrory, P. W. McKenty, D. D. Meyerhofer, J. F. Myatt, A. J. Schmitt, J. D. Sethian, R. W. Short, S. Skupsky, W. Theobald, W. L. Kruer, K. Tanaka, R. Betti, T. J. B. Collins, J. A. Delettrez, S. X. Hu, J. A. Marozas, A. V. Maximov, D. T. Michel, P. B. Radha, S. P. Regan, T. C. Sangster, W. Seka, A. A. Solodov, J. M. Soures, C. Stoeckl, and J. D. Zuegel, Phys. Plasmas **22**, 110501 (2015).

[4] R. Betti, M. Umansky, V. Lobatchev, V. N. Goncharov, and R. L. McCrory, Phys. Plasmas **8**, 5257 (2001); R. Betti, K. Anderson, V. N. Goncharov, R. L. McCrory, D. D. Meyerhofer, S. Skupsky, and R. P. J. Town, Phys. Plasmas **9**, 2277 (2002).

[5]M. Murakami, H. Nagatomo, T. Johzaki, T. Sakaiya, A. Velikovich, M. Karasik, S. Gus'kov, and N. Zmitrenko, Nucl. Fusion **54**, 054007 (2014).

[6]H. Azechi, T. Sakaiya, T. Watari, M. Karasik, H. Saito, K. Ohtani, K. Takeda, H. Hosoda, H. Shiraga, M. Nakai, K. Shigemori, S. Fujioka, M. Murakami, H. Nagatomo, T. Johzaki, J. Gardner, D. G. Colombant, J.W. Bates, A. L. Velikovich, Y. Aglitskiy, J. Weaver, S. Obenschain, S. Eliezer, R. Kodama, T. Norimatsu, H. Fujita, K. Mima, and H. Kan, Phys. Rev. Lett. **102**, 235002 (2009).

[7]Max Karasik, J. L. Weaver, Y. Aglitskiy, T. Watari, Y. Arikawa, T. Sakaiya, J. Oh, A. L. Velikovich, S. T. Zalesak, J. W. Bates, S. P. Obenschain, A. J. Schmitt, M. Murakami, and H. Azechi, Phys. Plasmas **17**, 056317 (2010).

[8]D. D. Ryutov, M. S. Derzon and M. K. Matzen, Rev. Mod. Phys. **72**, 167 (2000).

[9]M. G. Haines, Plasma Phys. Control. Fusion **53**, 093001 (2011).

[10]Y. Maron, A. Starobinets, V. I. Fisher, E. Kroupp, D. Osin, A. Fisher, C. Deeney, C. A. Coverdale, P. D. Lepell, E. P. Yu, C. Jennings, M. E. Cuneo, M. C. Herrmann, J. L. Porter, T. A. Mehlhorn, and J. P. Apruzese, Phys. Rev. Lett. **111**, 035001 (2013).

[11]C. A. Coverdale, C. Deeney, A. L. Velikovich, R. W. Clark, Y. K. Chong, J. Davis, J. Chittenden, C. L. Ruiz, G. W. Cooper, A. J. Nelson, J. Franklin, P. D. LePell, J. P. Apruzese, J. Levine, J. Banister, and N. Qi, Phys. Plasmas **14**, 022706 (2007); C. A. Coverdale, C. Deeney, A. L. Velikovich, J. Davis, R. W. Clark, Y. K. Chong, J. Chittenden, S. Chantrenne, C. L. Ruiz, G. W. Cooper, A. J. Nelson, J. Franklin, P. D. LePell, J. P. Apruzese, J. Levine, and J. Banister, Phys. Plasmas **14**, 056309 (2007).

[12]S. V. Lebedev, F. N. Beg, S. N. Bland, J. P. Chittenden, A. E. Dangor, M. G. Haines, K. H. Kwek, S. A. Pikuz, and T. A. Shelkovenko, Phys. Plasmas **8**, 3734 (2001); S. V. Lebedev, F. N.

Beg, S. N. Bland, J. P. Chittenden, A. E. Dangor, and M. G. Haines, Phys. Plasmas **9**, 2293 (2002).

[13]S. C. Bott, D. M. Haas, Y. Eshaq, U. Ueda, F. N. Beg, D. A. Hammer, B. Kusse, J. Greenly, T. A. Shelkovenko, S. A. Pikuz, I. C. Blesener, R. D. McBride, J. D. Douglass, K. Bell, P. Knapp, J. P. Chittenden, S. V. Lebedev, S. N. Bland, G. N. Hall, F. A. Suzuki Vidal, A. Marocchino, A. Harvey-Thomson, M. G. Haines, J. B. A. Palmer, A. Esaulov, and D. J. Ampleford, Phys. Plasmas **16**, 072701 (2009).

[14]K. Wu, Space Sci. Rev. **93**, 611 (2000).

[15]H.-Th. Janka, Annu. Rev. Nucl. Part. Sci. **62**, 407 (2012).

[16]N. Ohnishi, W. Iwakami, K. Kotake, S. Yamada, S. Fujioka and H. Takabe, J. Phys. Conf. Ser. **112**, 042018 (2008); T. Handy, T. Plewa, B. A. Remington, R. P. Drake, C. C. Kuranz, N. Ohnishi, and H. Takabe, High Energy Density Physics **8**, 331 (2012).

[17]J. C. Houck and R. A. Chevalier, Ap. J. **395**, 592 (1992); T. Foglizzo, P. Galletti, L. Scheck, and H.-Th. Janka, Ap. J. **654**, 1006 (2007); J. M. Laming, Ap. J. **659**, 1449 (2007).

[18]L. D. Landau and E. M. Lifshitz, *Fluid Mechanics*, 2nd English edition (Pergamon Press, Oxford, 1987).

[19]J. G. Wouchuk and J. Lopez Cavada, Phys. Rev. E **70**, 046303 (2004).

[20]J. W. Bates, Phys. Rev. E **91**, 013014 (2015).

[21]L. I. Sedov, Doklady Akademii Nauk SSSR **52**(1), 17 (1946) (in Russian); L. I. Sedov, *Similarity and Dimensional Methods in Mechanics* (Pergamon, New York, 1960).

[22]E. T. Vishniac, Ap. J. **274**, 152 (1983); D. Ryu and E. T. Vishniac, Astrophys. J. **313**, 820 (1987).

[23]V. M. Ktitorov, Atomic Science and Technology Issues, Ser. Theoretical and Applied Physics, No. 2, p. 28 (1984) (in Russian); V. M. Ktitorov, *Stability of diverging shock waves*, presented at the 8[th] International Workshop on the Physics of Compressible Turbulent Mixing, December 9-14, 2001, California Institute of Technology, Pasadena, California,

http://www.damtp.cam.ac.uk/iwpctm9/proceedings/IWPCTM8/slides/t15/t15.pdf

[24]J. Grun, J. Stamper, C. Manka, J. Resnick, R. Burris, J. Crawford, and B. H. Ripin, Phys. Rev. Lett. **66**, 2738 (1991).

[25]M. J. Edwards, A. J. MacKinnon, J. Zweiback, K. Shigemori, D. Ryutov, A.M. Rubenchik, K. A. Keilty, E. Liang, B. A. Remington, and T. Ditmire, Phys. Rev. Lett. **87**, 085004 (2001).

[26]J. M. Laming and J. Grun, Phys. Rev. Lett. **89**, 125002 (2002).

[27]A. L Velikovich, S. T. Zalesak, N. Metzler, and J. G. Wouchuk, Phys. Rev. E **72**, 046306 (2005).

[28]A. L. Velikovich, A. J. Schmit, N. Metzler, and J. H. Gardner, Phys. Plasmas **10**, 3270 (2003).

[29]Y. Aglitskiy, M. Karasik, A. L. Velikovich, V. Serlin, J. Weaver, T. J. Kessler, A. J. Schmitt, S. P. Obenschain, N. Metzler, and J. Oh, Phys. Rev. Lett. **109**, 085001 (2012).

[30]W. F. Noh, J. Comput. Phys. **72**, 78 (1987).

[31]E. P. Yu, A. L. Velikovich and Y. Maron, Phys. Plasmas **21**, 082703 (2014).

[32]A. L. Velikovich, J. L. Giuliani, S. T. Zalesak, J. W. Thornhill, and T. A. Gardiner, Phys. Plasmas **19**, 012707 (2012).

[33]J. L. Giuliani, A. L. Velikovich, S. T. Zalesak, P. Tzeferacos, D. Lamb, and J. W. Thornhill, *The Magnetized Noh Problem with Both Axial and Azimuthal Magnetic Fields*, presented at the The 41[st] IEEE International Conference on Plasma Science, Washington, DC, May 25 – 29, 2014.

[34]G. Guderley, Luftfahrt-Forsch. **19**, 302 (1942).

[35]K. P. Stanyukovich, *Unsteady Motion of Continuous Media* (Academic Press, New York, 1960); Ya. B. Zel'dovich and Yu. P. Raizer, *Physics of Shock Waves and High-Temperature Hydrodynamic Phenomena*, 2nd edition (Dover, New York, 2002).

[36]R. Lazarus, SIAM Journal of Numerical Analysis 18, 316 (1981); Errata, *ibid*., **19**, 1090 (1982).

[37]M. Murakami, J. Sanz, and Y. Iwamoto, Phys. Plasmas **22**, 072703 (2015).

[38]The supplemental material presenting the formulas and derivations necessary to make the text of the paper, together with the references, fully self-contained, is available from the authors upon request, murakami-m@ile.osaka-u.ac.jp .

[39]A. L. Velikovich, J. G. Wouchuk, C. Huete Ruiz de Lira, N. Metzler, S. Zalesak, and A. J. Schmitt, Phys. Plasmas **14**, 072706 (2007).

[40]A. I. Kleev and A. L. Velikovich, Plasma Phys. Contr. Fusion **32**, 763 (1990).

[41]G. B. Whitham, *Linear and Nonlinear Waves* (John Wiley & Sons, New York, 1974).

[42]A. L. Velikovich, Phys. Fluids **8**, 1666 (1996).

[43]Y. Yang, Q. Zhang, and D. Sharp, Phys. Fluids **6**, 1856 (1994).

[44]J. G. Wouchuk, Phys. Rev. E **63**, 056303 (2001); Phys. Plasmas **8**, 2890 (2001).

[45]T. A. Gardiner and J. M. Stone, J. Comput. Phys. **205**, 509 (2005); J. M. Stone, T. A. Gardiner, P. Teuben, J. F. Hawley, and J. B. Simon, Astrophys. J., Suppl. **178**, 137 (2008).

[46]See https://trac.princeton.edu/Athena/ for the Athena Web site containing a code description, documentation, and a link for downloading a public version of the Athena code.

[47]J. M. Stone, personal communication.

[48]K. Kitamura and E. Shima, J. Comp. Phys. **245**, 62 (2013).